\documentclass[english,a4paper]{aa}
\usepackage[T1]{fontenc}
\usepackage[utf8]{inputenc}
\usepackage{color}
\usepackage{graphicx}
\usepackage{amssymb}

\makeatletter

\providecommand{\tabularnewline}{\\}
\newcommand{\lyxdot}{.}

\@ifundefined{definecolor}
 {\usepackage{color}}{}

\makeatletter

\makeatletter

\authorrunning{I. Saviane et al.}
\titlerunning{Spectrophotometry of the  PN in the Phoenix dwarf galaxy}

\@ifundefined{definecolor}
 {\usepackage{color}}{}

\makeatletter

\@ifundefined{definecolor}
 {\usepackage{color}}{}

\makeatletter

\usepackage{txfonts}

\@ifundefined{definecolor}
 {\usepackage{color}}{}

\providecommand{\\}{\\}

\newcommand{\ha}{H$\alpha$}
\newcommand{\hb}{H$\beta$}
\newcommand{\hg}{H$\gamma$}
\newcommand{\hd}{H$\delta$}
\newcommand{\hi}{H\,{\sc i}}
\newcommand{\te}{$T_{\mathrm e}$}
\newcommand{\de}{$n_{\mathrm e}$}
\newcommand{\teff}{$T_{\mathrm{eff}}$}
\newcommand{\tbb}{$T_{\mathrm{BB}}$}

\newcommand{\foiii}{[O~{\sc iii}]}

\newcommand{\foi}{[O~{\sc i}]}
\newcommand{\foii}{[O~{\sc ii}]}
\newcommand{\fsii}{[S~{\sc ii}]}
\newcommand{\fsiii}{[S~{\sc iii}]}

\newcommand{\fnii}{[N~{\sc ii}]}

\newcommand{\fariv}{[Ar~{\sc iv}]}

\newcommand{\fneiii}{[Ne~{\sc iii}]}
\newcommand{\fneiv}{[Ne~{\sc iv}]}

\newcommand{\ffeiii}{[Fe~{\sc iii}]}

\newcommand{\oii}{O~{\sc ii}}

\newcommand{\ciii}{C~{\sc iii}}
\newcommand{\civ}{C~{\sc iv}}
\newcommand{\fciii}{C~{\sc iii}]}

\newcommand{\fariii}{[Ar~{\sc iii}]}

\newcommand{\hei}{He~{\sc i}}
\newcommand{\heii}{He~{\sc ii}}

\newcommand{\cmt}{cm$^{-3}$}

\newcommand{\ffevi}{[Fe~{\sc vi}]}
\newcommand{\ffev}{[Fe~{\sc v}]}
\newcommand{\foiv}{[O~{\sc iv}]}

\def\sun{\hbox{$_\odot$}}

\makeatother

\makeatother

\makeatother

\makeatother

\makeatother

\usepackage{babel}

\begin{document}

\title{VLT/FORS1 spectrophotometry of the first planetary nebula discovered
in the Phoenix dwarf galaxy}

\author{Ivo Saviane\inst{1} \and Katrina Exter \inst{2} \and Yiannis
Tsamis \inst{3} \and Carmen Gallart \inst{4} \and Daniel Péquignot
\inst{5} \fnmsep%
\thanks{Based on data collected at the European Southern Observatory, VLT,
Chile, Proposal N. 076.D-0089(A) %
}}

\offprints{I.~Saviane}

\institute{ESO Chile, A. de Cordova 3107, Santiago \email{isaviane@eso.org}
\and {STScI, 3700 San Martin Dr, Baltimore, MD 21218 USA } \email{kexter@stsci.edu}
\and {Dept. of Physics and Astronomy, University College London,
Gower Str., London, UK} \email{ygt@star.ucl.ac.uk} \and {IAC,
c/ via Lactea s/n, La Laguna, Tenerife, Spain} \email{carme@iac.es}
\and {LUTH, Observatoire de Paris, CNRS, Université Paris Diderot;
5 Place Jules Janssen, 92190 Meudon, France}~\email{Daniel.Pequignot@obspm.fr}}

\date{Received xxx; accepted xxx}

\abstract{
A planetary nebula (PN) candidate was discovered during FORS imaging of
the Local Group dwarf galaxy Phoenix. }{ Use this PN to complement
abundances from red-giant stars. }{ FORS spectroscopy was used to
confirm the PN classification. Empirical methods and photoionization
modelling were used to derive elemental abundances from the emission
line fluxes and to characterize the central star. }{ For the elements
deemed most reliable for measuring the metallicity of the interstellar
medium (ISM) from which the PN formed, {{[}O/H]$\sim-0.46$ and
{[}Ar/H]$\sim-1.03$.} {[O/H] has lower measurement errors but
greater uncertainties due to the unresolved issue of oxygen enrichment
in the PN precursor star.}}
{
{Earlier than $2$\,Gyr ago (the lower limit of the
derived age for the central star) the ISM}
had Z = 0.002--0.008, a range slightly
more metal-rich than the one provided by stars. Comparing our
PN-to-stellar values to surveys for other dwarf Local Group galaxies,
Phoenix appears as an outlier.

}

\maketitle

\section{Introduction}

\subsection{The Phoenix dwarf and its metallicity}

Phoenix is a Local Group dwarf galaxy located $450$~kpc from the
Milky Way (M$_{V}=-10.1$, L$_{V}=9\times10^{5}$~L$_{\odot}$, M$=3.3\times10^{6}\, M_{\odot}$,
$(m-M)_{\circ}=23.2$; Mateo \cite{mateo98}). It has little current
star formation (Mart\'{\i}nez-Delgado, Gallart \& Aparicio \cite{martinezdelgado_etal99},
hereafter M99; Held, Saviane \& Momany \cite{held_etal99}, hereafter
H99) and little or no \ion{H}{i} gas (St-Germain et al. \cite{st-germain_etal99};
Gallart et al. \cite{gallart_etal01}; Irwin \& Tolstoy \cite{irwin_tolstoy02};
Young et al. \cite{youngetal2007}), which warrants its classification
as a transition-type dwarf galaxy. From the color and width of the
red-giant branch (RGB) one can infer quite a low metallicity for its
intermediate-age and old stars: {[}Fe/H]$=-1.85$, $\sigma$({[}Fe/H])$\simeq0.5$~dex
(H99). However, metallicity estimates from the position of the RGB
alone are quite uncertain in the case of composite stellar populations
due to the resulting age--metallicity degeneracy (see, e.g. Pont et
al. \cite{pont_etal04}). With a color-magnitude diagram (CMD) reaching
the oldest main-sequence turnoffs, the RGB age-metallicity degeneracy
may be partially broken since ages can be derived from the main-sequence,
and Z(t) can be estimated from the global fit of the distribution
of stars in the CMD using stellar evolution models. In this way, Hidalgo,
Aparicio \& Mart\'{\i}nez-Delgado-Delgado (\cite{hidalgo_etal07}; H07) have
found that the current metallicity should be about $Z=0.0015$.

In any case, spectroscopic data provide a more direct measurement
of the metallicity and metallicity distribution of the stars in a
galaxy. In the case of Phoenix, \ion{Ca}{ii} triplet spectra,
which provide global metallicities, could be obtained for the brightest
RGB stars in the galaxy. 
Measuring abundances directly, through high-resolution spectroscopy
of individual stars, is not feasible due to the large distance of
the galaxy. Moreover, there are no known \ion{H}{ii} regions,
so the only other opportunity to obtain information on the abundances
of particular elements is through a PN. During a FORS1 campaign (63.I-0642),
a PN candidate was identified near the center of the galaxy, and in
this paper we present our follow-up spectroscopy. 
The object was discovered due to its relative brightness in H$\alpha$
as compared with the $V$ band, together with its star-like appearance
(see Fig.~\ref{fig:Finding-chart-for}). 
Following current convention we name this object PN\,Phoenix\,J01:51:05.45$-$44:26:55.28
but, in a light-hearted spirit, we choose to call it Bennu (the ancient
Egyptian name for the phoenix).

\subsection{Chemical abundances from PNe}

PNe are the post-AGB (asymptotic giant branch) stage of low- to intermediate-mass
stars. Their chemical abundances are those of the envelope of the
precursor AGB star which was ejected via a {stellar}
wind. The composition of the nebulae are a mixture of elements affected
by the preceding nucleosynthesis and dredge-up cycles (e.g.\ He,
C, and N) and elements unadulterated by stellar evolution (e.g.\ Ar,
S, and in most cases also O and Ne). Measuring PN abundances is useful
since they complement abundance determinations from stars, and contribute
to the study of galactic chemical evolution (e.g., Shields \cite{shields02}).
For Phoenix, obtaining PN abundances is especially important, as this
is one of the most metal poor Local Group galaxies. While with only
one PN we are dealing with low number statistics, we note that the
scatter about the mean oxygen abundance for PNe (the most often considered
element in PN--galactic evolution studies) in other galaxies is usually
less than a factor of two (Stasińska \& Izotov \cite{stasinska_etal03}).
Lacking any other determination, a sole PN can still provide crucial
information.

\section{Observations and data reduction}

\begin{figure}
\begin{centering}
\includegraphics[width=1\columnwidth]{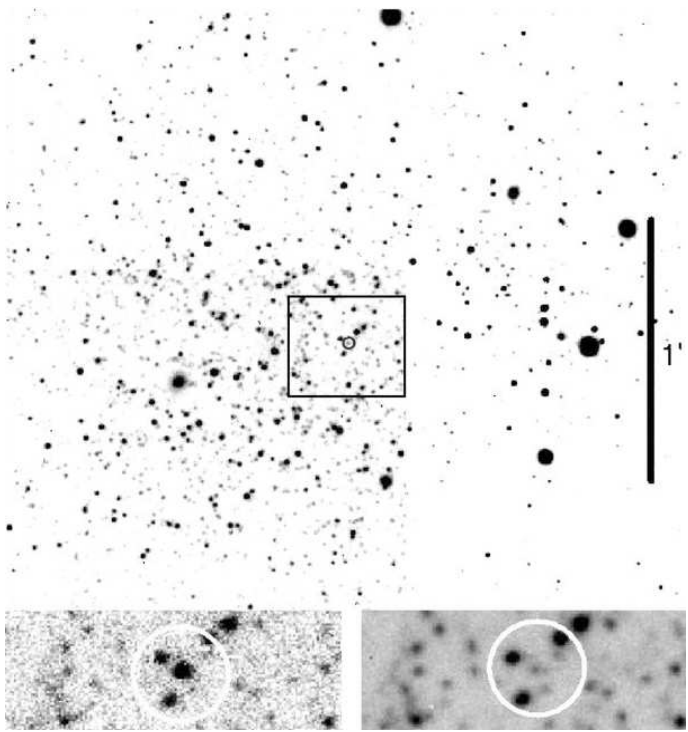}
\par\end{centering}

\caption{Finding chart for the Phoenix planetary nebula, obtained from FORS1
V-band imaging. The lower two insets show the appearance of the nebula
in the H$\alpha$ band (left) and V band (right). North is up and
east is to the left. \label{fig:Finding-chart-for}}

\end{figure}

\begin{figure*}
\noindent \begin{centering}
\includegraphics[width=0.7\textwidth,angle=-90]{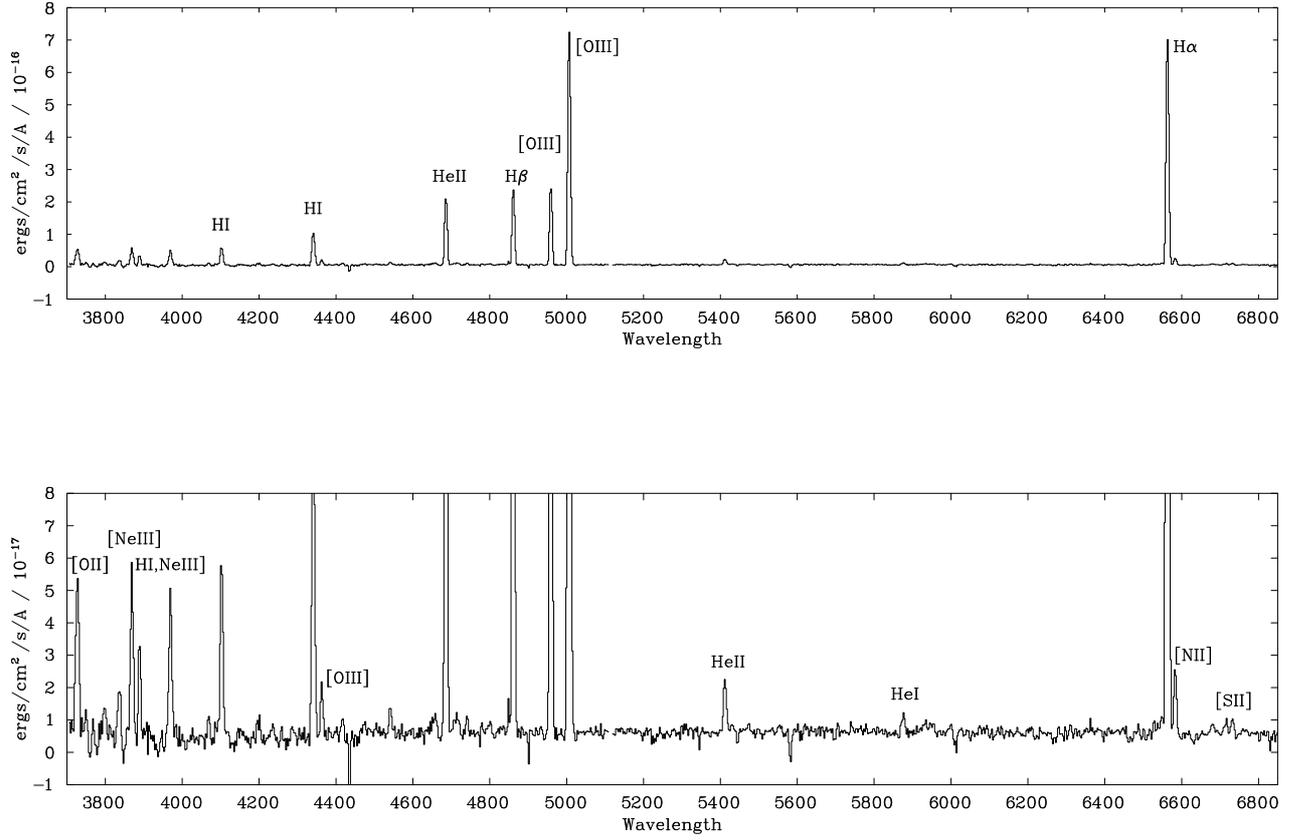}
\par\end{centering}

\caption{The calibrated and sky-subtracted spectrum of Bennu, above, and again
below, scaled to highlight the fainter lines. \label{fig:The-(not-dereddened)}}

\end{figure*}

The spectroscopic observations were carried out in service mode on
the night of November 26, 2005. The target can be identified in Fig.~\ref{fig:Finding-chart-for},
which shows part of one FORS2 frame taken in the V band. Three FORS1
spectra of $1730$\,s each were taken, with the spectrograph in long-slit
mode, a $0\farcs7$ slit, and with grism 300V+10. No order sorting
filter was used, in order to have a higher throughput in the blue
part of the spectrum, and the response function was then determined
by observing a blue standard star in the same configuration. Some
second-order contamination is expected at $\lambda>6600$~\AA, {but
it is less than $4\%$} for the {[}\ion{S}{ii}] lines at 6725\,\AA.
The nebula was observed at an airmass $\sim1.04$ and under a seeing
of $\sim1\arcsec$, while approaching and crossing the meridian. During
the evening twilight, the DA spectrophotometric standard EG21 from
Hamuy et al. (\cite{hamuy94}) was observed. {The
instrument mode was slightly different than the one used for the science
target: a $5\arcsec$ slit was created by placing the $19$ movable
slits of the MOS mode side by side, and aligning them along the parallactic
angle; moreover, only the central part of the CCD was read out. The
science data were instead taken with one of the fixed-width slit masks,
and the full CCD was read out.} {The two sets of}
calibration frames (biases, arc, and flat-fields) {for
the two configurations} were taken in the morning following the observations.

Reductions were carried out using a customized version of the EFOSC2
quick-look tool%
\footnote{http://www.ls.eso.org/%
}, which is based on the ESO-MIDAS data reduction system. The usual
bias subtraction and flat-fielding (from quartz lamps) were performed.
He+HgCd+Ar arc frames were used to compute the 2-D wavelength calibration.
The spectra were then linearly rebinned with a constant step of $\Delta\lambda=2.64$~\AA,
with a resulting wavelength coverage of 3707--8289\,\AA\ at a resolution
of $\sim8.2$\,\AA\ FWHM.

The two-dimensional sky frame was created by first sampling the sky spectrum in
two windows flanking the target spectrum, and then fitting the spatial gradient
with a one-degree polynomial. After sky-subtracting, the extraction window for
the emission-line spectrum was chosen based on the spatial profile of the
H$\alpha$ line, adjusting in order to reach an optimum signal-to-noise ratio
(SNR). The flux of the spectrophotometric standard star was instead summed over
almost the entire spatial profile, leaving out just the two sky windows. To
compute the response function, the instrumental spectrum was corrected for
atmospheric extinction, divided by the exposure time, and then divided by the
published spectrum, and the ratio was fit with a $12^{{\rm th}}$ degree
polynomial. {Note that the different instrument setup of the
standard does not affect the response function, which can therefore be applied
to the science spectra.} The three pre-processed, wavelength-calibrated,
sky-subtracted and extracted spectra of the PN were flux-calibrated using this
response function. Finally, the fiducial PN spectrum was obtained as a median
of the three individual exposures, and it is shown in
Fig.~\ref{fig:The-(not-dereddened)}. The typical SNR measured on the continuum
is five or better, while the {[}\ion{O}{iii}] $\lambda$4363 line was detected
with a SNR of $\sim24$. {The continuum is due in part to the
nebula and in part to the unresolved light of the Phoenix stellar population.
Photoionization models (see below) predict that $39\%$ of the observed
continuum intensity at 5000~\AA\ is due to the nebula. Because of the low
counts, no absorption lines can be detected, and no slope is visible. }

\section{Spectral measurements}

\begin{table}
\caption{Observed line fluxes. \label{tab:The-measured-fluxes}}

\begin{centering}
\begin{tabular}{lcccl}
\noalign{\vskip3pt} \noalign{\hrule} \noalign{\vskip3pt}
$\lambda$(\AA)  & $F$($\lambda$)$^{a}$  & $I$($\lambda$)$^{b}$  & Err.(\%)  & ID \tabularnewline
\noalign{\vskip3pt} \noalign{\hrule} \noalign{\vskip3pt}
3727  & 54.1  & 26.83  & 10  & \foii\tabularnewline
3749  & 10.4  & 5.73  & 50  & \hi\tabularnewline
3798  & 13.4  & 7.37  & 30  & \hi\tabularnewline
3836  & 18.6  & 10.20  & 15  & \hi\tabularnewline
3869  & 48.6  & 28.05  & 10  & \fneiii\tabularnewline
3889  & 24.5  & 13.54  & 10  & \hi, \hei\tabularnewline
3969  & 48.5  & 26.93  & 10  & \hi, \fneiii\tabularnewline
4069  & 6.43  & 3.66  & 25  & \fsii+$^{c}$\tabularnewline
4102  & 51.4  & 27.75  & 10  & \hi\tabularnewline
4341  & 95.1  & 50.43  & 10  & \hi\tabularnewline
4363  & 15.0  & 8.32  & 10  & \foiii\tabularnewline
4541  & 7.46  & 4.07  & 40  & \heii\tabularnewline
4657  & 5.59  & 3.02  & 50  & \civ, \ffeiii\tabularnewline
4686  & 182.  & 98.06  & 10  & \heii\tabularnewline
4714  & 10.1  & 5.44  & 30  & \fariv+$^{c}$\tabularnewline
4740  & 5.06  & 2.71  & 35  & \fariv\tabularnewline
4861  & 198.  & 100  & 10  & \hi\tabularnewline
4959  & 210.  & 110  & 5  & \foiii\tabularnewline
5007  & 651.  & 341  & 5  & \foiii\tabularnewline
5412  & 17.2  & 8.68  & 20  & \heii\tabularnewline
5876  & 5.70  & 2.77  & 30  & \hei\tabularnewline
6312  & 3.11  & 1.11  & 50  & \fsiii\tabularnewline
6563  & 630.  & 284  & 5  & \hi\tabularnewline
6583  & 19.9  & 9.37  & 15  & \fnii\tabularnewline
6717  & 3.57  & 1.67  & 50  & \fsii\tabularnewline
6731  & 5.61  & 2.62  & 50  & \fsii\tabularnewline
7136  & 5.95  & 2.73  & 25  & \fariii\tabularnewline
\hline
 &  &  &  & \tabularnewline
\end{tabular}
\par\end{centering}

\begin{description}
\item {[{$^{a}$}] In units of 10$^{-18}$ erg\,cm$^{-2}$\,s$^{-1}$.}

\item {[{$^{b}$}] De-reddened and ratioed to \hb\ = 100. All \hi\ lines de-blended
from \heii; \foii\ 3727 de-blended from \fsiii\ and two \hi\ lines.}

\item {[{$^{c}$}] Blends: \fsii\ 4069 + \ffev\ 4071 + \ciii\ 4069;
\fariv\ 4711 + \hei\ 4713 + \fneiv\ 4714+16.}

\end{description}

\end{table}

Emission line fluxes and centroids were measured with the ELF Gaussian
fitting routines of DIPSO (Howarth et al. \cite{howarth_etal98})
and independently checked with the fitting routines of the MIDAS package.
The FWHM for faint lines were fixed from near-by bright lines and
were a free parameter for unresolved blends. The fluxes are reported
in Table\,\ref{tab:The-measured-fluxes}. The measured blended \hb$+$He\,\textsc{ii}
flux is 1.98$\times10^{-16}$\,erg\,cm$^{-2}$\,s$^{-1}$ and the
adopted de-reddened and de-blended \hb\ intensity%
\footnote{Correction for the He\,\textsc{ii} contribution is calculated from
theoretical He\,\textsc{ii} line ratios to the uncontaminated He\,\textsc{ii}\,4686\,\AA\ flux.%
} is 3.50$\times10^{-16}$\,erg\,cm$^{-2}$\,s$^{-1}$.
{This value, which is used in the models of Section 4.2, incorporates
a correction of $+$28\% as the employed slit had a width of 0.$''$7
and the PN was observed under $\sim$1$''$ seeing.}

The continuum in the very blue and red is fairly noisy, and a conservative
approach was taken to deciding between a true line and a noise spike. This is
especially important for the He\,\textsc{i} lines -- because of the dependence
of the calculated oxygen abundance on the helium abundances when using the
`ICF' method (see below) -- and the {[}S\,\textsc{ii}] ratio, which is used to
calculate the electron density. 
{
The sky is noisy in the [S\,\textsc{ii}] region, and the
[S\,\textsc{ii}] fluxes very low. To estimate the effect this has on the
measurement of the [S\,\textsc{ii}] ratio 
we performed a sky subtraction in the standard manner (using the sky
spectrum extracted from the slit and resulting in the fluxes quoted in
Table~1) and a sky subtraction using rather a polynomial fit to the sky
in this spectral region. The [S\,\textsc{ii}] ratio from the poly-fit
result is up to $70\,\%$ greater that from the true-sky subtraction.}
Regarding He$^{+}$, only the He\,\textsc{i} $\lambda$5876 line,
de-blended from He\,\textsc{ii} $\lambda$5869.02 (5--29), is considered
in our analysis as the 4471\,\AA\ and 6678\,\AA\ lines are upper
limit detections.

{
The \ha/\hb\ Balmer line ratio was used to calculate the extinction constant,
yielding a value of c(\hb)$=$0.16. The He\,\textsc{ii} Pickering series is a
contaminant of the H\,\textsc{i} lines in high excitation nebulae such as this
one and their contribution was estimated via their theoretical ratios to
He\,\textsc{ii} $\lambda$4686 following Storey \& Hummer
(\cite{storey_hummer95}); 13.5\%\ of the $\lambda$4686 flux at H$\alpha$ is due
to He\,\textsc{ii} Pi (4--6) and 5.1 per cent at H$\beta$ is due to Pi (4--8),
under case B conditions. The intrinsic value of the H\,\textsc{i} ratio to
compare to when computing c(\hb) depends on the electron temperature, \te, and
density, \de; we iterated once between determining \te\ and \de\ before
computing c(\hb). As the He\,\textsc{ii} $\lambda\lambda$4541.59, 5411.53 lines
are well detected, we also computed, to compare, the reddening using the
He\,\textsc{ii} $\lambda$4686/$\lambda$5411 and $\lambda$5411/$\lambda$4542
ratios; this yielded a weighted average of $0.19\pm50$\%. We adopted the \hi\
Balmer value to deredden the observed line fluxes, and used the Galactic
extinction law of Howarth (1983).
}

A 16\%\ correction has been applied to the {[}O~\textsc{ii}] $\lambda$3727
doublet which is blended with the H\,14 $\lambda$3721.9, {[}S~\textsc{iii}]
$\lambda$3721.6 and H\,13 $\lambda$3734.4 lines: the {[}S~\textsc{iii}]
$\lambda$3721 flux was estimated from a comparison with the de-reddened flux of
{[}S~\textsc{iii}] $\lambda$6312.1, which originates from the same upper level.
The latter line is blended with He~\textsc{ii} $\lambda$6310.8 (5--16) in high
excitation PNe and that flux was retrieved via its theoretical ratio relative
to He~\textsc{ii} $\lambda$4686. The \emph{I}(H\,13)/\emph{I}(H$\beta$) and
\emph{I}(H\,14)/\emph{I}(H$\beta$) intensity ratios were estimated using
theoretical H\,\textsc{i} line emissivities as above.

In Table~\ref{tab:The-measured-fluxes} suggested ion identifications are given.
{The bluest \hi\ line ratios, at $\lambda\le3968$, are a factor of $1.3$--$1.8$
in excess of that expected theoretically, suggesting that the adopted blue
continuum, possibly affected by interstellar absorption lines, is too low. {
\hi\,$\lambda$3771 is not detected in the spectrum.}}

\section{Chemical abundances and central star parameters}

\subsection{Empirical analysis}

{
A determination of chemical abundances was first carried out using the
semi-empirical ICF (ionization correction factor) scheme reported in
Kingsburgh \& Barlow (1994). The abundances of each observed ionic
species relative to H$^{+}$ are added up and multiplied by their
corresponding ICF to yield elemental abundances relative to
hydrogen. Input values in these calculations are the plasma \te\ and
\de, whose values were computed from the de-reddened {[}O\,\textsc{iii}]
$\lambda$4363/($\lambda$5007+$\lambda$4959) and {[}S\,\textsc{ii}]
$\lambda$6717/6731 ratios respectively.  }

\begin{table}
\caption{ {Empirical ionic abundances$^{a}$ and ICF values for two electron
densities.}}

\begin{centering}
\begin{tabular}{lcc|lcc}
\noalign{\vskip3pt} Ion  & 500  & 3700 & Ion  & 500  & 3700 \tabularnewline
\hline
\foii\  & 17.1  & 25.7 & \fsii\  & 0.38  & 0.54 \tabularnewline
\foiii\  & 280.  & 279. & \fsiii\  & 3.87  & 4.30 \tabularnewline
icf(O)  & 3.96  & 3.96 & icf(S)  & 1.82  & 1.61 \tabularnewline
\fnii\  & 5.42  & 5.62 & \fariii\  & 0.93  & 0.93 \tabularnewline
icf(N)  & 73.1  & 50.0 & \fariv\  & 1.82  & 1.64 \tabularnewline
\fneiii\  & 57.4  & 57.4 & icf(Ar)  & 1.01  & 1.02 \tabularnewline
icf(Ne)  & 4.18  & 4.30 &  &  & \tabularnewline
\hline
\end{tabular}
\par\end{centering}

\begin{description}
\item { [{$^{a}$}] In $10^{-7}$ by number relative to H.}
\end{description}

\end{table}

\begin{table}
\caption{{Abundances of Bennu vs solar$^{\mathrm{a}}$\label{tab:Empirically-calculated-abundances}}}

\begin{centering}
\begin{tabular}{lcccc}
\noalign{\vskip3pt} Element & \de= 500 & = 3700  & Model  & Sun \tabularnewline
\hline
He/H                  & .125(20) & .125(20) & .108(10) & .098 \tabularnewline
O/H\,$\times$10$^{4}$ & 1.18(14) & 1.21(17) & 1.7(13)  & 4.9 \tabularnewline
N/O\,$\times$10       & 3.4(24)  & 2.3(26)  & 1.5(15)  & 1.6 \tabularnewline
Ne/O\,$\times$10      & 2.0(30)  & 2.0(26)  & 1.4(15)  & 1.8 \tabularnewline
S/O\,$\times$10$^{3}$ & 6.5(50)  & 6.5(50)  & 10.$^{\mathrm{b}}$ & 37. \tabularnewline
Ar/O$\times$10$^{3}$  & 2.4(30)  & 2.2(30)  & 2.3(25)  & 8.5 \tabularnewline
\hline
\end{tabular}
\par\end{centering}

\begin{description}
\item 
{[{$^{\mathrm{a}}$}] Proto-solar abundances from Asplund et
al. (\cite{asplund_etal05}), except for O/H from Allende-Prieto et al.\
(\cite{allende}). Percentage errors in parenthesis. 
}

\item 
{[{$^{\mathrm{b}}$}] S/Ar solar; S/O\,$\times$10$^{3}$ may as well be 20
according to model outputs.
}

\end{description}

\end{table}

The same atomic data references as in Exter et al.\
(\cite{exter_etal04}) were used, with the exception of using helium
where effective recombination coefficients were taken from Smits
(\cite{smits96}) for Case B and adopting corrections for collisional
excitation contributions as in Benjamin, Skillman and Smits
(\cite{benjaminetal}).

{For the density measured from the {[}S\,\textsc{ii}] ratio: to show the
effect that measurement uncertainties (in particular, sky subtraction)
has on the derived density we carried out here calculations using the
{[}S\,\textsc{ii}] ratio measured from the two sky-subtraction methods
mentioned in Sec.~3.  A higher density 
comes from the
lower ratio (from the fluxes we adopted and list in Table~1) and a lower
density 
from the higher ratio. 
The effect on the Te value is minor,
\te$=16\,600\pm1000$\,K, while \de\ changes from 500 to 3700\,cm$^{-3}$. 
{In Tables\,2 and 3 we give the ionic and total abundances with this
range of $n_{\rm e}$ and $T_{\rm e}$ values.}
{
The percentage 1$\sigma$ errors given in parenthesis in Table\,3 include the
flux and \te\ measurement errors and not those inherent to the empirical
methods (see Exter et al. \cite{exter_etal04}). In Table 3, errors for the
model abundances (see Sect.~4.2) correspond either to the scatter among several
model fits or to the flux errors.
}

{
Since the \heii\ flux is very high, the nebula cannot be radiation bounded.
Then the formulation of Kaler \& Jacoby (\cite{kaler_jacoby89}) 
for estimaging the \teff\ of the central star only provides
an upper limit to
\teff\ ($<$ 322\,kK) and the one of 
Zijlstra \& Pottasch (\cite{zijlstra_pottasch89}) a lower limit to
$L_{{\rm bol}}$ ($>$ 270\,L\sun). The method of Zanstra
(\cite{1927ApJ....65...50Z}) to obtain a
\teff(\heii) requires knowledge of the stellar continuum (see below), while the
energy balance method of Stoy (\cite{1933MNRAS..93..588S}) requires
knowledge of important gas coolants outside the optical
range. Self-consistent photoionization modelling based on the optical
line spectrum is the only available method for the determination of
\teff\ and $L_{{\rm bol}}$.  }

\begin{figure}
\includegraphics[width=1\columnwidth]{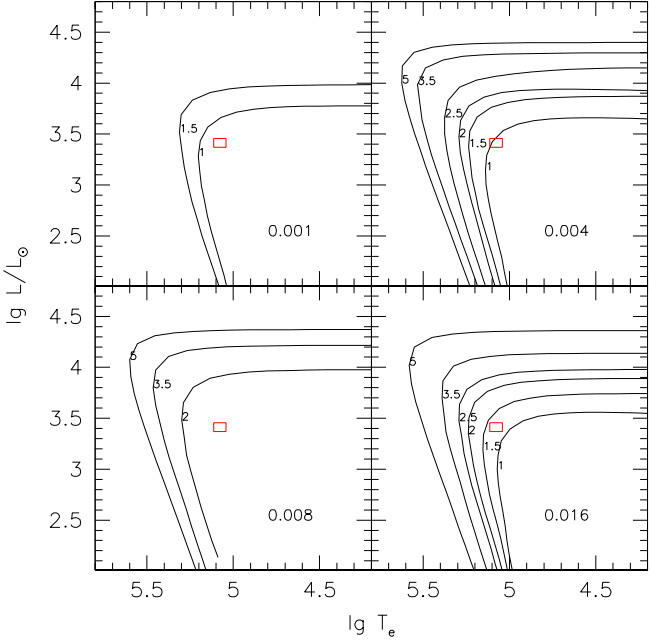}

\caption{The theoretical, H-burning WD tracks from VW94 are represented here
in the $\log L/L_{\odot}$ vs. $\log T_{{\rm e}}$ plane. The four
panels show four different metallicities, identified by the label,
and for each metallicity tracks of different initial mass are plotted.
A label near the knee of each track identifies its initial mass, in
solar units. The open box represents the permitted range for the central
star of Bennu, with parameters determined via photoionization modeling
of the nebula. \label{fig:The-theoretical,-H-burning}}

\end{figure}

\begin{figure}
\includegraphics[width=1\columnwidth]{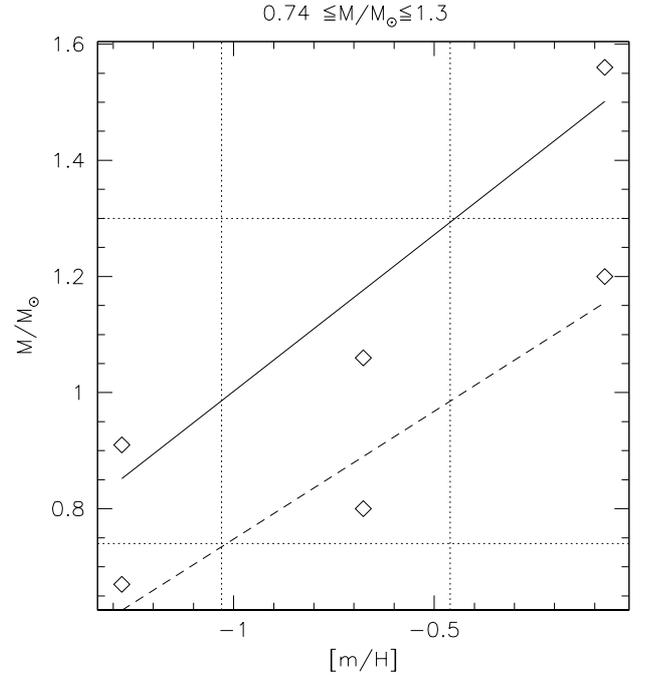}

\caption{The dependence of the progenitor mass on the metallicity of the best-fitting
WD track is illustrated by this plot. The lines are least-square fits
to the discrete points corresponding to 
{\bf three of}
the four metallicities available
in VW94 tracks (see Fig.\,3). The fits have been repeated for the
two combinations of $(T_{{\rm e}}-\Delta T_{{\rm e}};L-\Delta L)$
(dashed line) and $(T_{{\rm e}}+\Delta T_{{\rm e}};L+\Delta L)$ (solid
line) allowed by our error estimates. The vertical dotted lines show
the permitted metallicity range determined from the nebular emission
lines, and the horizontal dotted lines mark the limits of the permitted
progenitor mass. \label{fig:The-dependence-of-progenitor-mass} }

\end{figure}

\subsection{Photoionization modelling \label{sub:model}}

{
Photoionization models were computed with the code \textsc{nebu} (e.g.,
P\'equignot and Tsamis \cite{pequignot_tsamis05}), which was also used in a
comprehensive study of the Sagittarius dwarf galaxy PN population (Dudziak et
al. \cite{dudziak_etal00}; {P\'equignot et al. \cite{pequignot_etal00}}; Zijlstra
et al. \cite{zijlstra_etal06}). 
{Computations were done assuming that both the
gas filling factor and the covering factor (solid angle as seen from the
central star) were unity.} The actual geometry of the nebula is not
known due to lack of imaging and the inner radius $R_{{\rm in}}$ is a
first free parameter.  A two-sector model is suggested by the
coexistence of a very strong \heii\ $\lambda$4686 line with \foii\ and
\fsii\ lines which implies that even though the nebula is strongly
matter-bounded along most radial directions, some directions must be
radiation-bounded to a large degree (One-sector models are however
considered in Appendix~\ref{sec:photo-model}).
The premise of two-sector models comprising
both optically thick and thin components (advocated by, e.g., Clegg et
al. \cite{1987ApJ...314..551C} in their analysis of the typical Galactic
PN NGC\,3918) is a realistic one, given that the majority of nearby
spatially-resolved PNe are replete with optically thick inhomogeneities
in the form of clumps/filaments/torii embedded in a more tenuous
medium. The \fsii\ doublet ratio can be used to constrain the density
or, more conveniently, the thermal pressure $P_{{\rm out}}$ of the
(peripheral) thick clumps.
The \fariv\ doublet ratio 
is neither suitably accurate nor
sensitive enough to 
\de\ to
constrain the inner high-ionization region density, $N_{{\rm in}}$(H)
(equivalently, $P_{{\rm in}}$), which is therefore a second free parameter.
Adopting a sufficiently flexible description for the gas distribution allows
the exploration of a realistically large space of solutions. In practice, the
pair of (\te, \de) is obtained at each point by solving iteratively the
statistical balance equations, assuming a variable gas pressure, $P$, given
here as a function of the radial optical depth, $\tau$, at 13.6 eV:
}

\begin{equation}
P(\tau)=\frac{P_{{\rm out}}+P_{{\rm in}}}{2}+\frac{P_{{\rm out}}-P_{{\rm in}}}{\pi}\tan^{-1}\Bigl[\kappa\log\Bigl(\frac{\tau}{\tau_{{\rm c}}}\Bigl)\Bigr].
\end{equation}
{At the first step of the computation ($\tau$ $=$ 0) the pressure is $P_{{\rm
in}}$, while at the last step ($\tau$ $=$ $\infty$; in practice $\gg$1) it is
$P_{{\rm out}}$. The requirement for a sector (Sector~2) comprising
radiation-bounded clumps immersed in a dilute medium dictates that $P_{{\rm
in}}$ $<$ $P_{{\rm out}}$. The third parameter $\tau_{c}$ at which
$P$($\tau_{{\rm c}}$) = ($P_{{\rm in}}$ $+$ $P_{{\rm out}}$)/2 corresponds to
the inner boundary of the generic clump. The fixed fourth parameter, $\kappa$
$=$ 30, is large enough to ensure a sharp clump boundary. For consistency, the
optical depth $\tau_{1}$ of the dilute matter-bounded Sector~1 must be
substantially less than $\tau_{{\rm c}}$ (in practice the condition $\tau_{{\rm
c}}$/$\tau_{1}$ $\geq$ 1.2 was adopted).
}

{
The ionizing spectrum is described as a black body of temperature \tbb\ and
luminosity $L$. According to stellar atmosphere models for hot stars, the
`colour temperature' of the bulk of the photoionizing continuum (the best
equivalent black-body temperature) is
larger than the actual effective temperature \teff. 
When black-body spectra are used instead of genuine stellar atmosphere models
in computing photoionization models, a recommended rough correction, adopted
here, is:
}

\begin{equation}
T_{\mathrm{eff}} = T_{\mathrm{BB}} - 15\,\textrm{kK}.
\end{equation}

\noindent
(see, e.g., Dudziak et al. \cite{dudziak_etal00}).

\begin{table}
\caption[]{{Constraints on model parameters.}}
\label{tab_mod_param}
\begin{tabular}{ll}
\noalign{\vskip3pt} \noalign{\hrule} \noalign{\vskip3pt}
Parameter  & Constraint \tabularnewline
\noalign{\vskip3pt} \noalign{\hrule} \noalign{\vskip3pt}
$R_{{\rm in}}$\, \ \ \ \ \ free & standard 1$\times$10$^{17}$\,cm \tabularnewline
$N_{{\rm in}}$(H) free & $P_{{\rm in}}$ $<$ $P_{{\rm out}}$; stdd 1200\,\cmt; `\fariv\ $\lambda$4740/$\lambda$4711+' \tabularnewline
\tbb\, \ \ \ \ free & $\tau_{{\rm c}}$/$\tau_{1}$ $>$ 1.2; `upper limit \foi\ $\lambda$6300' \tabularnewline
\noalign{\vskip3pt} \noalign{\hrule} \noalign{\vskip3pt}
$\tau_{1}$(thin)& absolute flux $I$(\hb) \tabularnewline
$f_{2}^{{\rm cov}}$(thick) & \foii\ $\lambda$3727; ($f_{1}^{{\rm cov}}$ + $f_{2}^{{\rm cov}}$ $=$ 1) \tabularnewline
$P_{{\rm out}}$  & ratio \fsii\ $\lambda$6731/$\lambda$6716 \tabularnewline
$\tau_{{\rm c}}$/$\tau_{1}$ & \hei\ $\lambda$5876 (see text) \tabularnewline
$L$     & ratio \fariv\ $\lambda$4740 / \fariii\ $\lambda$7135 \tabularnewline
\noalign{\vskip3pt} \noalign{\hrule} \noalign{\vskip3pt}
He      & \heii\   $\lambda$4686 \tabularnewline
C  & ratio \foiii\ $\lambda$4363/($\lambda$4959+$\lambda$5007) \tabularnewline
N       & \fnii\   $\lambda$6584 \tabularnewline
O       & \foiii\  $\lambda$5007+$\lambda$4959 \tabularnewline
Ne      & \fneiii\ $\lambda$3869 \tabularnewline
S       & S/Ar = 4.37 by number (solar); `\fsiii\ $\lambda$6312' \tabularnewline
Ar      & \fariii\ $\lambda$7135 + \fariv\ $\lambda$4740 \tabularnewline
Fe & Fe/H = 4$\times$10$^{-7}$ by number; `upper limit \ffevi' \tabularnewline
\noalign{\vskip3pt} \noalign{\hrule} \noalign{\vskip3pt}  & \tabularnewline
\end{tabular}
\end{table}

{
The model parameters and corresponding dominant observational and technical
constraints are listed in Table~\ref{tab_mod_param}. Some potential constraints
which were dismissed on account of the large uncertainties they entail are
given in inverted commas. Thus, we assumed that S/Ar was solar (4.37 by number;
Lodders \cite{2003ApJ...591.1220L}). Also, no iron lines were detected and Fe/H was given a constant
arbitrary value. There are 11 ($+$2 implicit for S and Fe) spectroscopic
constraints controlling as many model parameters, leaving \tbb\ as a third free
parameter, in addition to $R_{{\rm in}}$ and $N_{{\rm in}}$(H) (see above). Gas
cooling is dominated by carbon and oxygen line emission. For lack of measured
carbon lines (no UV coverage), the carbon abundance is constrained by the
energy balance through the \foiii\ ratio temperature. Given the three free
parameters and assuming that all other constraints are systematically
fulfilled, for increasing $L$, the computed \fariv/\fariii\ increases and \hei\
decreases moderately, while, for increasing $\tau_{{\rm c}}$/$\tau_{1}$,
\fariv/\fariii\ increases slowly and \hei\ decreases again, until a limit is
reached asymptotically for $\tau_{{\rm c}}$/$\tau_{1}$ $>$ 2.5--3. Thus $L$ and
$\tau_{{\rm c}}$/$\tau_{1}$ are primarily controlled by the ionization balances
of argon and helium respectively. Other correspondences between parameters and
constraints in Table~\ref{tab_mod_param} are straightforward.
}

{
The domain of `exact' solutions according to the criteria listed in
Table~\ref{tab_mod_param} has been scanned. Many iterations were needed to
fulfill all constraints simultaneously and not all combinations of the 3 free
parameters led to a solution. Results are summarized in
Table~\ref{tab_mod_range}. Given $R_{{\rm in}}$ = 1$\times$10$^{17}$\,cm (our
`standard' value), the full range of \tbb\ leading to a solution was determined
for the 3 given values of $N_{{\rm in}}$(H) (\cmt) = 900 ($P_{{\rm
out}}$/$P_{{\rm in}}$ = 1.35), 1200 (standard, $P_{{\rm out}}$/$P_{{\rm in}}$ =
1.95), and 1600 ($P_{{\rm out}}$/$P_{{\rm in}}$ = 2.4). The ratio
\fariv/\fariii\ being accounted for (correct choice of $L$), a minimum \tbb\ is
obtained when \hei\ happens to be computed too large and a maximum \tbb\
is obtained when $\tau_{{\rm c}}$/$\tau_{1}$ decreases down to
1.2. After several trials, the resulting accessible domain on the
(\tbb/kK, $L$/10$^{36}$\,erg\,s$^{-1}$) plane was found to be an
elongated ellipse, whose long-axis boundaries were at (120, 11.3) and
(150, 8.7) respectively for $N_{{\rm in}}$(H)~$\sim$~800 (the minimum
value since $P_{{\rm out}}$/$P_{{\rm in}}$~$\sim$~1.2), and
$\sim$~1700\,\cmt: the larger the $N_{{\rm in}}$(H), the more compact
the nebula and the smaller the corresponding $L$ are. Thus, for a given
$R_{{\rm in}}$, $L$ correlates inversely with \tbb. The `1$\sigma$ box'
for the standard case, not including observational uncertainties, is
\tbb/kK $=$ 135$\pm$10 and $L$/10$^{36}$\,erg\,s$^{-1}$ $=$
10$\pm$1. For $N_{{\rm in}}$(H) = 1200, `exact' models with $R_{{\rm
in}}$/10$^{17}$\,cm $=$ 1.4 instead of 1.0 were also obtained for a few
\tbb's, showing that derived parameter values were about similar to
those obtained previously, except for a 10\% upward shift of $L$
(Col.~5, Table~\ref{tab_mod_range}). Conversely, for $R_{{\rm
in}}$/10$^{17}$\,cm $<$ 1.0, the previous solutions for standard
$R_{{\rm in}}$ essentially apply. Model outputs are displayed in
Table~\ref{tab_mod_appendix} of Appendix~\ref{sec:photo-model},
including one solution
(D135, Col.~3) belonging to the domain of most favorable values of model
parameters (Table~\ref{tab_mod_range}) and examples of models which are
in some way unsatisfactory, including one-sector models.
A more thorough discussion of the uncertainties is also given in
Appendix~\ref{sec:more-discussion}.
}

\begin{table}
\caption{{Range of model parameters for `exact' solutions \label{tab_mod_range}}}

\begin{tabular}{lccccc}
\noalign{\vskip3pt}
Parameter            &        \multicolumn{4}{c}{}                & 1-$\sigma$ box \tabularnewline
\hline
$R_{{\rm in}}$ (10$^{17}$\,cm) & 1.0 &  1.0    &    1.0   &    1.4     &      --     \tabularnewline
$N_{{\rm in}}$(H) (\cmt) &    900  &   1200    &   1600   &   1200     &      --     \tabularnewline
\hline
\tbb\ (kK)           & 127$\pm$5 & 135$\pm$7  & 145$\pm$3 & 135$\pm$7  & 135$\pm$10 \tabularnewline
\teff\ (kK)          & 112$\pm$5 & 120$\pm$7  & 130$\pm$3 & 120$\pm$7  & 120$\pm$10 \tabularnewline
$L$ (10$^{36}$\,erg\,s$^{-1}$) & 11.$\pm$.1 & 10.$\pm$.1 & 9.$\pm$.1 & 11.$\pm$.1 & 10.$\pm$1. \tabularnewline
He/H$\times$10$^{3}$ & 108$\pm$0 & 108$\pm$0  & 109$\pm$1 & 108$\pm$0  & 108$\pm$1  \tabularnewline
C/H$\times$10$^{6}$  & 197$\pm$1 & 212$\pm$4  & 230$\pm$4 & 222$\pm$4  & 214$\pm$17 \tabularnewline
O/H$\times$10$^{6}$  & 165$\pm$7 & 172$\pm$9  & 178$\pm$4 & 169$\pm$9  & 170$\pm$12 \tabularnewline
N/O$\times$10$^{3}$  & 149$\pm$4 & 147$\pm$1  & 150$\pm$1 & 147$\pm$1  & 149$\pm$3  \tabularnewline
Ne/O$\times$10$^{3}$ & 142$\pm$3 & 143$\pm$1  & 153$\pm$2 & 149$\pm$1  & 145$\pm$5  \tabularnewline
Ar/O$\times$10$^{5}$ & 232$\pm$6 & 228$\pm$5  & 226$\pm$2 & 228$\pm$5  & 230$\pm$7 \tabularnewline
\hline
\end{tabular}

\end{table}

{
Model elemental
abundances are relatively stable within the domain of `exact' solutions
(Col.~6, Table~\ref{tab_mod_range}). Model outputs (see Col.~3 of
Table~\ref{tab_mod_appendix}) suggest that any value of S/Ar between solar
(adopted) and twice solar is possible. For higher \teff's, O/H can be somewhat
larger and Ar/O slightly smaller (Col.~4, Table~\ref{tab_mod_appendix}).
}

\subsection{Parent star}

{
For the central star, in the previous section we obtained best values
\teff~$= 120\pm10$\,kK and $L_{{\rm bol}} = 2600\pm260$\,L$_\odot$.} 
To determine an age and mass for Bennu we have compared these model
estimates to the H-burning white-dwarf (or PN central star) evolutionary
tracks of Vassiliadis \& Wood (\cite{vassiliadis_wood94}; VW94), as
shown by Fig.~\ref{fig:The-theoretical,-H-burning}.
{As the metallicity increases, the progenitor mass of
the best-fitting track increases.}
To have a quantitative estimate, for each metallicity the tracks were
interpolated at the ($\log$ \teff, $\log$ $L$) position of the WD, thus
obtaining a mass-metallicity relation, which is shown in
Fig.~\ref{fig:The-dependence-of-progenitor-mass}.  {Due to the small
number of tracks and the large extrapolation from the PN location to any
of them, the $Z=0.008$ point was omitted from the fit.} The procedure
was repeated for the two extremes of the $(\log T_{{\rm e}};\log L)$
combinations allowed by the errors on the two quantities. By fitting the
discrete points with linear regressions, one has continuous
mass--metallicity relations, so entering with a metallicity value an
estimate of the progenitor mass can be obtained.  If we adopt the widest
metallicity range discussed below
{(${\rm [m/H]=-1.03}$ from $\rm [Ar/H]$ to ${\rm [m/H]=-0.46}$ from $\rm [O/H]$)}, then the range
of progenitor mass is between
{ $0.74$ and $1.30\,{\rm M}_{\odot}$}.

This low mass chimes well with the N/O ratio and He/H nebular abundance,
both being sufficiently low to indicate a non-Type\,I, low central
star mass status (according to e.g.\ the criterion of Kingsburgh
\& Barlow, \cite{kingsburgh_barlow94}). From the VW94 tracks, the
nebular age would be 
{23\,000\,yrs} 
at the lower mass estimate,
and 
{7\,000\,yrs} 
at the higher. To convert the progenitor
mass range to a main-sequence (MS) age range, we looked at the isochrones
of Pietrinferni et al. (\cite{pietrinferni_etal04}), for two values
of the metallicity $Z=0.002$ and $Z=0.008$ ({[}m/H]$=-1$ and $-0.4$,
respectively).

The lower mass value {is just below the minimum isochrone mass 
($0.77 \, \rm M_{\odot}$), so the star might be as old as the Universe.}
To convert the
higher mass value into an age, one needs to decide between canonical
core convection, or convection with overshooting, which yields larger
ages. In the case of canonical convection, one obtains 
{$2.0$
or $2.7$ Gyr} 
for the two metallicities, while tracks with overshooting
give 
{$3.3$ and $4.0$ Gyr}, 
respectively. Thus the overall
permitted age range is {greater than 2~Gyr}.

As discussed below, it is also quite possible that the metallicity given
by Ar (and S) is closer to the real PN value. In that case, if we take the
{model argon abundance, then}
{${\rm [m/H]}=-1.03\pm{0.1}$},
and by repeating the exercise above, the resulting mass range is
{
$0.70$--$1.04$~$M_{\odot}$.  }
Taking only isochrones at $Z=0.002$, the
mass interval is then converted into an age interval from 
{$4.7$~Gyr to the age of the Universe.}

\section{Discussion \label{sec:Discussion}}

\subsection{Oxygen as a metallicity tracer? \label{sub:Oxygen-as-a}}

The gas-phase elemental abundance in nebulae which is the easiest
to compare to stellar values is that of oxygen (i.e.\ {[}O/H]); {[}O\,\textsc{iii}]
lines are collectively strong and usually contain most of the flux
from oxygen ions (except in cases of extremely metal-poor nebulae),
and thus measurement and abundance-determination errors are small.
Along with sulphur, argon and neon, oxygen is generally considered
to be unaffected by endogenous processes within the progenitor stars.
For instance, when comparing the oxygen abundances in PNe to those
of \ion{H}{ii} regions in a given galaxy the former are expected
to be always lower (or equal) than the latter; H\,\textsc{ii} regions
after all represent the present-day metallicity of the ISM, whereas
PNe originate from older stellar populations when galactic metallicities
are likely to have been smaller. But this is not always found to be
the case.

A number of studies of PNe in low-metallicity environments have pointed
to a possible enhancement of O/H ascribed to `self-{enrichment}'
processes occurring during the 3$^{{\rm rd}}$ dredge-up phase. PN
vs. H\,\textsc{ii} region abundance comparisons reveal a positive
difference between the two at low O/H values (at about log\,O/H$<-4$;
see e.g.\ Peña et al. \cite{pena_etal07}; Richer \& McCall
\cite{richer_mccall07}).  For NGC 3109 (Peña 2007) this is about 0.3
dex, and Kniazev et al.\ (\cite{kniazev_etal07}) adopted a similar value
of 0.27\,dex to `correct' their PN oxygen abundance for
`self-{enrichment}'. Comparison between models and observations of PNe in
the Magellanic Clouds (Leisy \& Dennefeld \cite{leisy_dennefeld}) have
also suggested that oxygen in PNe can be enhanced relative to the
precursor metallicities, particularly for the lower mass progenitors.
These studies have indicated that argon or sulphur should be used
instead of oxygen, as an indicator of the metallicity of the ISM from
which the PNe formed, with argon preferred as having the more secure
spectral determination; Péquignot et al. (\cite{pequignot_etal00}) and
Péquignot \& Tsamis (\cite{pequignot_tsamis05}) reached similar
conclusions. 

It is uncertain whether Bennu is likely to contain
endogenous oxygen as there are no H\,\textsc{ii} regions in Phoenix
to compare to. Its O/H is however higher (by $>0.3$\,dex) than the
range of values found in those PNe for which an oxygen enhancement
has been suggested (although see Zijlstra et al.\ \cite{zijlstra_etal06}).
In addition,
following Kniazev et al.\ (\cite{kniazev_etal07})
and based on the nebula's Ne/O, S/O and Ar/O ratios, we should apply
no correction for the presence of endogenous oxygen. However, the
{[}Ar/H] abundance is lower than {[}O/H], so if we adopted the former
as an indicator of the metallicity of the parent ISM, as suggested
by Leisy \& Dennefeld, then we should indeed conclude that endogenous
oxygen is present in the nebula. Since with our present knowledge
we have no exact way to decide upon this issue, in the rest of the
discussion we will adopt a metallicity for the Phoenix galaxy within
the range given by {[}O/H] and {[}Ar/H], and with the understanding
that the more metal-poor end is favoured.

\subsection{The age--metallicity relation of Phoenix}

{
In the previous sections we concluded that the spectral properties
of the newly discovered PN in Phoenix are consistent with a progenitor
mass of 
$1.0\pm0.3\, M_{\odot}$ 
and a metallicity 
${\rm [m/H]=-0.75\pm0.29}$~dex
(or $Z=0.002$ to $Z=0.008$), corresponding to an age of 
$7.9 \pm 5.9$~Gyr.
Or if argon and sulfur abundances are favored over oxygen, then the
progenitor mass and age range are 
$0.87 \pm 0.17 \, \rm M_{\odot}$ 
and 
$9.2 \pm 4.5$~Gyr,
for 
${\rm [m/H]}=-1.03\pm{0.1}$.
}

Since this is the first direct spectroscopic determination of this
galaxy's metallicity, it also represents the first solid constraint
to its age--metallicity relation (AMR). It is therefore interesting
to see how this fits within our present knowledge of the chemical
evolution of Phoenix and its star-formation history (SFH).

The evidence accumulated since its discovery by Schuster \& West (\cite{schuster_west76})
and its recognition as a dwarf galaxy by Canterna \& Flower (\cite{canterna_flower77}),
clearly shows that Phoenix has had an extended star formation. The
presence of an old, globular-cluster like population in Phoenix was
first established by the discovery of an extended horizontal branch
(HB) at $V\sim23.8$ (H99; M99). The metallicity of the intermediate--old
populations was estimated from the color of the RGB by van de Rydt
et al. (\cite{vdry+91}) and H99, as {[}Fe/H]$=-2.0$~dex and {[}Fe/H]=$-1.81\pm0.10$~dex,
respectively. A significant spread in RGB color was also found which,
if it was only due to a metallicity range, would correspond to a dispersion
of about $0.5$ dex. A very low metallicity at an age of $\sim13$~Gyr
is then the first point in the AMR of Phoenix.

A number of stars above the RGB tip were interpreted as AGB stars
by H99. They are representatives of an intermediate age population
($3$ to $10$ Gyr), which was estimated to be about $30$--$40\%$
of the total stellar population of the galaxy. The metallicity of
these stars could not be estimated, but given the RGB width, it could
be even up to $-0.3$~dex, if we adopted an upper limit of $3\sigma$
(i.e. $-1.8+3\times0.5$~dex).

The presence of a young stellar population was established quite early
(Canterna \& Flower \cite{canterna_flower77}; Ortolani \& Gratton
\cite{ortolani_gratton88}), and together with the possible association
of an \ion{H}{i} cloud, lead to a morphological classification
for the Phoenix as intermediate between dSph and dIrr galaxies (e.g.,
Young et al. 2007; Carignan, Demers, \& Côté, \cite{carignan_etal91}).
According to H99 the star formation episode started at least $0.6$~Gyr
ago, and lasted until $10^{8}$~yr ago. Since the galaxy has no \ion{H}{ii}
regions, and since its main-sequence stars are too faint, no spectroscopic
determination exists for the current metallicity of Phoenix.

The current metallicity and future evolution of Phoenix depend on the
presence of gas associated with the galaxy. The recent measurement of
the optical radial velocity of Phoenix (Irwin \& Tolstoy
\cite{irwin_tolstoy02}) has shown that it is the same as that of `cloud
A' of Young and Lo (\cite{young_lo97}), confirming a long-suspected
association (St-Germain et al. \cite{st-germain_etal99}). This means
that the \ion{H}{i} mass is $\sim10^{5}M_{\odot}$, and as shown in H99,
it can be explained by mass lost from RGB and AGB stars, and PNe, over
the last $\sim2$\ Gyr.  {And according to Young et al. (2007) the
gas will not be able to escape the galaxy, so SF might start again in
the future.}
{However currently
SF from
molecular gas is inhibited,
since no CO was detected by Buyle et
al. (\cite{buyle_etal06}), and Jackson et al. (\cite{jackson_etal06})
found no diffuse 8~$\mu$m emission from dust.
The (unknown) metallicity of Phoenix youngest stars therefore represents
the present end point of the galaxy's chemical evolution.}

It is also important to note that age gradients exist throughout the
galaxy, with all young stars concentrated in an inner component, while
the old population is more extended and oriented N--S (M99; H99).
The fraction of intermediate-age stars over old stars also decreases
going from the central to the outer regions (Hidalgo et al. 2007).
Moreover, additional support for an extended SFH is given by the study
of the variable-star content of Gallart et al. (\cite{gallart_etal04}).
They find both anomalous Cepheids and short-period classical Cepheids,
which can be explained if the metallicity has been low ({[}Fe/H]$=-1.3$)
for most of the galaxy's lifetime.

The star formation history for the central part of Phoenix has been
derived by Holtzman et al. (\cite{holtzman_etal00}; H00) and Young
et al. (2007), using a WFPC2 CMD reaching the oldest main sequence
turnoffs. Hidalgo et al. (2007) present a determination of both the
star formation rate as a function of time (SFR(t)) and the metallicity
as a function of time (Z(t)) for the same field and an outer one.
All these authors agree on that Phoenix has had an almost continuous
SFH, and with a roughly constant star formation rate, somewhat decreasing
toward the present time. The star formation rate at intermediate ages
($\simeq$ 6--2 Gyr) seems to have been somewhat larger than immediately
before and after, specially in the central part of the galaxy where
Bennu resides. Integrating their age distribution one gets that $15\%$
of the stars have ages between $1.5$ and $3$~Gyr, and another $15\%$
have ages between $3$ and $10$~Gyr, which is consistent with the
fraction of intermediate-age stars estimated by H99 (see above). Although
H00 do not give a real AMR, from their `population box' one can see
that the {[}Fe/H] increases from $\sim-2$ at the earliest stage up
to $-$(1.5--1) at $\sim3$~Gyr, then reaches $\sim-1$--0 at $\sim1$~Gyr
and stays constant thereafter. In the case of H07, they predict quite
low metallicities up to the current epoch. The maximum is reached
ca. $1$~Gyr ago, and it is less than $-1$ in ${\rm [m/H]}$.

Turning now back to Bennu, we can place its representative point in the
AMR relations of H00 and H07, which are the only two studies making
quantitative derivations. In the age range of Bennu's progenitor, most
of the simulated stars of H00 have metallicities around ${\rm
[m/H]\sim-1}$, but there is also a smaller group of objects with
metallicities centered around ${\rm [m/H]=-0.5}$. Adopting instead the
age range suggested by argon and sulfur abundances, the metallicities of
H00 are around ${\rm [m/H]\sim-1.5}$ with a tail up to ${\rm
[m/H]=-1}$. Therefore the metallicity we measure tends to be higher than
that of H00 in the same age range, but still compatible within the
uncertainties. A metallicity lower than ours is also predicted by the
simulations of H07, whose AMR generally stays more metal-poor than that
of H00. As it was recalled above, even the peak of the relation happens
at a metallicity lower than that of the PN. While an independent study
would be needed to resolve this issue, we can conclude that existing
AMRs predict, for the age of Bennu, a metallicity that is lower than the
one we measure, particularly so for the H07 study.

\subsection{Phoenix in the context of Local Group galaxies }

\subsubsection{The number--luminosity relation}

A recent review of the PN population in Local Group (LG) galaxies can be found
in Corradi \& Magrini (\cite{corradi_magrini06}; CM06), where they list $20$
galaxies containing at least $1$ nebula. These objects belong mostly to the
northern sky, where the most systematic searches are on-going. There is a very
good linear correlation between the $V$ luminosity of a galaxy and the number
of PN candidates, which predicts that below L$_{V}=10^{7}$\,L$_{\odot}$ the
probability of finding a PN is very low. This means that in principle Phoenix,
which has a L$_{V}\simeq10^{6}$\,L$_{\odot}$, should have no PNe. However the
number of nebulae also depends on the SFH of a galaxy, and in particular an
enhancement at intermediate ages can also increase the number of detected PNe
{(see Buzzoni et al. \cite{buzzoni_etal06} for a thorough
treatment of the luminosity-specific frequency of PNe)}. Some evidence of this
possibility is given in CM06, where it is shown that galaxies with enhanced SFH
in the age interval $2$ to $8$~Gyr have indeed a larger number of PNe (for
example NGC~205 and the SMC). The presence of a PN in Phoenix despite its low
luminosity could therefore be a sign of substantial SFH at intermediate ages.
This is nicely consistent with the large number of AGB stars detected by H99,
who, as recalled above, estimated that $\sim37\pm12$\% of the total population
should consist of stars with ages of $3$ to $10$ Gyr. The enhanced SFR between
$2$ and $3$~Gyr found by H00 (for a central field very close to Bennu) is also
in line with this finding. An additional possibility, discussed in the next
section, is that the luminosity of Phoenix is underestimated, either because of
measurement uncertainties, or because of mass being stripped from the galaxy by
the tidal field of the Milky Way. A combination of these two effects might be
able to produce a luminosity which is a factor of $10$ larger than what is
currently published, and so to reconcile Phoenix with the number-luminosity
relation.

\subsubsection{The luminosity-metallicity relation}

\begin{figure}
\noindent \begin{centering}
\includegraphics[width=1\columnwidth]{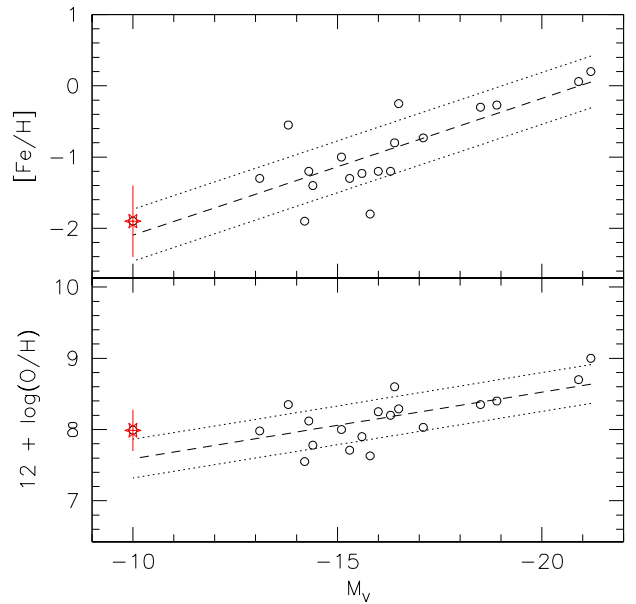}
\par\end{centering}

\caption{The luminosity-metallicity relation for Local Group galaxies, with
data from Corradi (priv. comm.); the {[}Fe/H] values come from stellar
results (old--intermediate populations) and {[}O/H] is from nebular
abundances (PNe and H\,\textsc{ii} regions; intermediate--young populations).
M$_{{\rm V}}$ were taken from Mateo (\cite{mateo98}). The dashed
line represents an unweighted linear fit, while the dotted lines represent
the $\pm1\sigma$ dispersion around the average line. \label{fig:The-L-Z-relation}}

\end{figure}

In Fig.~\ref{fig:The-L-Z-relation} we plot the iron and oxygen abundances
vs. the absolute $V$ luminosity for a sample of Local Group galaxies
(R. Corradi, priv. comm.). The asterisk represents Phoenix, with ${\rm [Fe/H]}$
from H99 and $12+\log{\rm (O/H)}$ from this paper. Its error bars
are the 1-sigma dispersion for the iron abundance given by the spread
in color of the RGB, while for the oxygen abundance the bar represents
our permitted range. The luminosity has been computed by taking the
apparent luminosity $V=13.2$ from de Vaucouleurs \& Longo (\cite{devc_longo88})
and converting it into $M_{V}=-10\pm0.13$ with the distance modulus
of H99 based of the HB luminosity: $(m-M)_{0}=23.21\pm0.08$. An error
of $0.1$~mag was assumed for the catalog luminosity. Since the relation
is steeper for {[}Fe/H], the figure might be telling us that lower-mass
galaxies undergo a comparatively larger metal enrichment than higher-mass
galaxies, when going from intermediate--old populations to younger
populations (note that the range of the $y$-axis is the same in both
panels). In a simple closed-box picture of chemical evolution, this
would mean that smaller galaxies have been able to convert larger
fractions of their gas into stars. However the figure might also be
telling us that {[}O/Fe] is enhanced in low-mass galaxies, as discussed
in Sect.~\ref{sub:Oxygen-as-a}. The scatter of the relations is
probably due to a combination of measurement errors, intrinsic abundance
scatter, and age differences of the metallicity tracers. Indeed old
to intermediate stars contribute to the ${\rm [Fe/H]}$ values estimated
with the color of the RGB, and intermediate to young populations contribute
to the ${\rm [O/H]}$ values from PNe and \ion{H}{ii} regions.

Figure \ref{fig:The-L-Z-relation} shows that the value of ${\rm [O/H]}$
of Phoenix is more than 1-sigma above the average, for its luminosity.
There are several possible explanations for this fact. First, it is
possible that the PN's oxygen abundance would lead us to overestimate,
as discussed in Sect.~\ref{sub:Oxygen-as-a}, the metallicity of
the galaxy. {Indeed recent theoretical AGB models
by Marigo (\cite{marigo01}; M01) predict positive oxygen yields for
$0.8\lesssim M/M_{\odot}\lesssim3.5$, with a strong dependence on
metallicity. This is due to the fact that, as metallicity decreases,
the thermal-pulse AGB phase increases its duration. Therefore it allows
more dredge-up episodes, which also happen to be more efficient. As
we found above, the progenitor star of Bennu is in the range $1\lesssim M/M_{\odot}\lesssim1.6$
and so it is formally in the range where oxygen enhancement is expected.
However in M01 models the yield has a maximum for masses between $2\, M_{\odot}$
and $3\, M_{\odot}$ and it becomes very modest below $\sim1.3M_{\odot}$.
On the other hand the minimum metallicity in M01 models is ${\rm [m/H]=-0.7}$,
and we might expect a larger oxygen enhancement if the metallicity
of Bennu is closer to $-1$ as suggested by the argon abundance. Lacking
a custom theoretical model constructed for the Phoenix PN progenitor,
the conclusion is that an endogenous production of oxygen might explain
the position of Phoenix in the luminosity-metallicity relation (see
also Magrini et al. \cite{magrini_etal05} for an extensive discussion
of oxygen enhancement in PNe). Note also that,} taking argon as a
metallicity indicator (i.e. the lowest point of the error bar in the
figure), the abundance would be within the r.m.s. dispersion of the
relation.

{It is also} possible that the progenitor was born
within an ISM recently enriched by Type~II SNe. But perhaps the problem
lies in the luminosity of the galaxy. The $V$ luminosity is flagged
as `Low quality data' in the \emph{NASA/IPAC Extragalactic Database},
and for example the $R$ luminosity from Lauberts and Valentijn (\cite{lauberts_valentijn89})
is $1$ magnitude brighter. A luminosity of $\sim-11$ would bring
the galaxy within the 1-sigma dispersion of the relation. And finally
another possibility is that the current mass of the galaxy is not
representative of the mass at the time when the PN was born. In fact
M99 find that the radial profile of Phoenix can be well fit by a King
model, which suggests a tidal truncation by the gravitational field
of the Milky Way. It also means that part of the galaxy mass was lost
due to dynamical evolution, and hence that the luminosity of Phoenix
was larger in the past.

\section{Summary and conclusions}

A planetary nebula was recently discovered in the Phoenix dwarf galaxy,
the first ever found in this stellar system. In this paper we presented
our follow-up spectroscopic data obtained at the ESO/VLT with FORS1.
With a total exposure time of $1.4$ hours we measured emission line
fluxes down to $9\times10^{-19}$~erg~cm$^{-2}$~s$^{-1}$. Such
high sensitivity allowed to detect, together with the Balmer series,
all lines of oxygen, neon, sulfur, helium, argon, and nitrogen above
the quoted threshold. To calculate the abundances of these elements,
both the empirical method, and photoionization modeling were employed.
The element-to-hydrogen abundance ratios were found to be consistent
with each other for the two methods, but the second method yields
more reasonable luminosity and temperature for the central star. The
oxygen abundance is larger than that of argon and sulfur. {This
might be due to endogenous production of oxygen in the progenitor,
or it might be caused by a truly higher oxygen abundance of the ISM
where the progenitor was born. To decide between the two hypotheses
an independent measurement of the abundance would be needed, which
however does not exist.} Therefore we based our discussion on a metallicity
range comprised between 
{[}m/H]$=-1.03$ 
(sulfur and argon abundances)
and {[}m/H]$=-0.46$ 
(oxygen abundance). Our conclusions would be essentially
the same if we restricted the metallicity range to that allowed by
argon and sulfur alone.

{
Using VW04 tracks and Pietrinferni et al. (\cite{pietrinferni_etal04})
isochrones, we found that the progenitor star of Bennu should have
an age comprised between 
$2.0$ 
and 
$13.7$ 
Gyr. The more restrictive
low metallicity would give an age 
$9.2\pm 4.5$~Gyr. 
}
So despite a
considerable uncertainty, the Phoenix PN allows to put a constrain
on the galaxy's age-metallicity relation. This shows that existing
AMRs underestimate the metallicity at intermediate ages, by as much
as $\sim0.6$~dex, even when adopting the most restricted {[}m/H]
range.

Within the general picture of LG dwarf galaxies, the presence of a
PN in Phoenix is unexpected given the galaxy's low luminosity (Corradi
\& Magrini \cite{corradi_magrini06}). Moreover, an extrapolation
of a linear L-Z relation to the luminosity of Phoenix would suggest
that the metallicity of the galaxy is larger than that of other LG
galaxies of comparable luminosity (at an intermediate age). On the
other hand there are no other galaxies at such low luminosity with
measured nebular oxygen abundance, so it might be possible that the
relation simply deviates from linearity for the lowest-mass galaxies.
Another possibility is that the problem resides in the luminosity.
The galaxy might be more luminous than quoted in the literature, so
a modern measurement of the integrated light would be very desirable.
It is also very likely that Phoenix lost a fraction of its mass through
tidal interaction with the gravitational potential of the Milky Way
(M99), so its representative point in the L-Z relation should me moved
to higher values of L. This would also help with the PN-number vs.
luminosity relation, thereby explaining the presence of Bennu.

In conclusion, this work demonstrates that the detection of even a
single PN in a dwarf galaxy, can provide essential information on
its chemical and dynamical evolution, and it can lead to a better
understanding of its past star formation history and mass build-up.

\begin{acknowledgements}
David
 Mart\'{\i}nez-Delgado-Delgado identified the candidate PN during run 63.I-0642
by blinking the $V$ and $R$ images. This research has made use of
the NASA/IPAC Extragalactic Database (NED) which is operated by the
Jet Propulsion Laboratory, California Institute of Technology, under
contract with the National Aeronautics and Space Administration. {We
thank the referee Romano Corradi for a meticulous examination of our
paper which lead to a substantially improved manuscript.}
\end{acknowledgements}

\appendix

\section{Photoionization model outputs}

\label{sec:photo-model}

{
Two-sector models (`D') using standard assumptions (Table~\ref{tab_mod_param})
and one-sector models (`S'), labelled by \tbb\ in kK, are presented in
Table~\ref{tab_mod_appendix}. All but one (D135) fall outside the domain where
`exact' solutions can be found (Sect.~\ref{sub:model},
Table~\ref{tab_mod_range}), and `best fit' values are adopted for \hei,
\fariii, \fariv\ and even \heii. D105 cannot account for \heii\ for any He/H
and $\tau_{1}$, and the computed \hei\ is unacceptably large. D160 illustrates
the maximum \tbb\ obtained when the computed \hei\ $\lambda$5876 intensity is
allowed to depart from observation by at most 33\%. For even larger \tbb, \hei\
is underestimated and the ratio \fariv/\fariii\ is not accounted for either.
Also, the optically thick sector can develop a warm neutral zone beyond the
ionization front that emits (unobserved) \foi\ $\lambda$6300.
}

{
Computed \fsii\ and \fsiii\ intensities for D135 suggest that S/Ar may be up to
twice larger than the adopted solar value, but this is not judged significant,
given the weakness of observed \fsiii\ $\lambda$6312. From the strongest
predicted iron line, \ffevi\ $\lambda$5145, an upper limit to Fe/H is twice the
adopted value, so that Fe/Ar is less than 0.25 solar: as in usual PNe, most of
the iron is likely to be locked into dust grains. In
Table~\ref{tab_mod_appendix}, the ratio \fariv\ $\lambda$4711+/$\lambda$4740
increases with \tbb\ due to the growing contribution of \fneiv.
}

{
For single-sector uniformly matter-bounded models to simultaneously account for
\heii\ and \foii\ requires, at the same time, a high \tbb, a low $L$ and small
ionization parameter (large $R_{{\rm in}}$ and/or large $N_{{\rm in}}$). These
models typically fail to reproduce the argon ionization balance,
underestimating the \fariv/\fariii\ ratio by factors of $\gtrsim$3, and imply
large He/H (in the models displayed, He/H is too small). These models are
rejected, as well as their low value of O/H. Any positive detection of \foi\
would further invalidate the single-sector assumption.
}

\begin{table}

\caption{{Model parameters and results. \label{tab_mod_appendix}}}

\begin{tabular}{lrccccc}
\noalign{\vskip3pt} \noalign{\hrule} \noalign{\vskip3pt}
\multicolumn{2}{l}{Model$^a$}  & D105  & D135  & D160  & S180  & S310  \tabularnewline
\noalign{\vskip3pt} \noalign{\hrule}\noalign{\vskip3pt}
\multicolumn{2}{l}{$L$ (L$_{\odot}$)}  & 2423  & 2550  & 2420  & 2080 & 2084  \tabularnewline
\noalign{\vskip2pt} 
\noalign{\vskip3pt} \multicolumn{2}{l}{$R_{\rm in}$ (10$^{17}$\,cm)} & 1.00 & 1.00 & 1.00 & 2.58 & 1.82\tabularnewline
\multicolumn{2}{l}{$R_{{\rm out}}$(thin) (10$^{17}$\,cm)}  & 2.08  & 2.15  & 2.14  & 2.97 & 2.06\tabularnewline
\multicolumn{2}{l}{$R_{{\rm out}}$(thick) (10$^{17}$\,cm)} & 2.91  & 2.98  & 2.80  &  --  &  -- \tabularnewline
\multicolumn{2}{l}{$f^{{\rm cov}}$(thick)}     & 0.027 & 0.013 & 0.012 & --   & --  \tabularnewline
\multicolumn{2}{l}{$N_{{\rm in}}$(H) (\cmt)} & 1200  & 1200  & 1200  & 1000  & 1600\tabularnewline
\multicolumn{2}{l}{$<$\de$>$ (\cmt)}  & 1942  & 1860  & 1870  & 1590 & 2940 \tabularnewline
\multicolumn{2}{l}{$<$$T$(H$^+$)$>$ (10$^4$\,K)} & 1.752  & 1.780  & 1.781 & 1.730 & 1.669 \tabularnewline
\multicolumn{2}{l}{$\tau_{{\rm c}}$}         & 1.00  & 1.75  & 1.65  & 0.90  & 0.90\tabularnewline
\multicolumn{2}{l}{$\tau$(thin) (13.6\,eV)} & 0.50  & 0.79  & 1.09  & 1.37  & 3.40\tabularnewline
\multicolumn{2}{l}{\hb(thin) fraction}      & 0.73  & 0.87  & 0.89  & 1.00  & 1.00 \tabularnewline
\multicolumn{2}{l}{$M_{{\rm ion}}$ (10$^{-2}$\,M$_{\odot}$)} & 6.83 & 7.02 & 6.85 & 6.11 & 3.37\tabularnewline
\noalign{\vskip2pt} \multicolumn{2}{l}{He/H} & 0.122 & 0.108 & 0.105 & 0.120 & 0.110\tabularnewline
\multicolumn{2}{l}{C/H ($\times$10$^{-5}$)} & 13.0  & 22.0  & 26.5  & 11.7  & 25.0\tabularnewline
\multicolumn{2}{l}{N/H ($\times$10$^{-5}$)} & 1.30  & 2.56  & 3.10  & 1.42  & 1.24\tabularnewline
\multicolumn{2}{l}{O/H ($\times$10$^{-5}$)} & 8.35  & 17.5  & 22.0  & 7.72  & 8.26\tabularnewline
\multicolumn{2}{l}{Ne/H ($\times$10$^{-5}$)} & 1.22 & 2.57  & 3.49  & 1.29  & 1.50\tabularnewline
\multicolumn{2}{l}{S/H ($\times$10$^{-7}$)} & 14.9  & 17.3  & 19.7  & 12.7  & 11.4\tabularnewline
\multicolumn{2}{l}{Ar/H ($\times$10$^{-7}$)} & 3.40 & 3.96  & 4.55  & 2.90  & 2.60\tabularnewline
\multicolumn{2}{l}{Fe/H ($\times$10$^{-7}$)} & 4.00 & 4.00  & 4.00  & 4.00  & 4.00\tabularnewline
\noalign{\vskip3pt} \textit{Lines}  & $I_{{\rm obs}}$$^{b}$  & \multicolumn{5}{c}{\textit{Predicted intensity / Observed intensity}}\tabularnewline
\noalign{\vskip3pt}
\ha\ 6563   & 284.  & 0.99  & 0.99  & 1.00  & 1.04  & 1.06\tabularnewline
\hg\ 4340   & 50.4  & 0.94  & 0.94  & 0.94  & 0.94  & 0.94\tabularnewline
\hd\ 4101   & 27.7  & 0.96  & 0.97  & 0.97  & 1.00  & 1.01\tabularnewline
\hei\ 5876   & 2.8  & 2.56  & 1.00  & 0.75  & 1.13  & 0.74\tabularnewline
\heii\ 4686  & 98.1 & 0.88  & 1.00  & 1.00  & 0.98  & 0.91\tabularnewline
\fciii\ 1908 & --  & (591)  & (573)  & (582) & (625) & (1267)\tabularnewline
\civ\ 1549   & --  & (1239) & (1864) & (1978) & (667) & (835)\tabularnewline
\civ\ 4658$^{c}$ & 3.0: & 0.12 & 0.18 & 0.22 & 0.12  & 0.09\tabularnewline
\fnii\ 6584   & 9.4  & 1.00  & 1.00  & 1.00  & 1.00  & 1.00\tabularnewline
\foi\ 6300   & $<$2. & 0.64  & 0.74  & 0.77  & 0.03  & 0.03\tabularnewline
\foii\ 3727   & 26.8 & 1.00  & 1.00  & 1.00  & 1.00  & 1.00\tabularnewline
\foiii\ 4363  & 8.3  & 1.00  & 1.00  & 1.00  & 1.00  & 1.00\tabularnewline
\foiii\ 5007+ & 451. & 1.00  & 1.00  & 1.00  & 1.00  & 1.00\tabularnewline
\foiv\ 26$\mu$m & -- & (229) & (551) & (619) & (151) & (107)\tabularnewline
\fneiii\ 3868 & 28.0 & 1.00  & 1.00  & 1.00  & 1.00  & 1.00\tabularnewline
\fsii\ 4069$^{c}$ &3.7: & 0.31 & 0.22 & 0.24 & 0.25  & 0.21\tabularnewline
\fsii\ 6716  & 1.7:  & 0.60  & 0.40  & 0.39  & 0.50  & 0.41\tabularnewline
\fsii\ 6731  & 2.6:  & 0.63  & 0.41  & 0.40  & 0.49  & 0.41\tabularnewline
\fsiii\ 6312  & 1.1: & 0.94  & 0.71  & 0.74  & 1.29  & 1.11\tabularnewline
\fariii\ 7136  & 2.7 & 1.38  & 1.00  & 1.00  & 1.63  & 1.45\tabularnewline
\fariv\,4711$^{c}$ &5.4 & 0.76 & 0.95 & 1.02 & 0.41  & 0.29\tabularnewline
\fariv\ 4740  & 2.7  & 0.93  & 1.00  & 1.00  & 0.47  & 0.34\tabularnewline
\ffevi\ 5145  & $<$3. & 0.33 & 0.48  & 0.42  & 0.26  & 0.20\tabularnewline
\noalign{\vskip3pt} \noalign{\hrule}\noalign{\vskip3pt} &&&&&&\tabularnewline
\end{tabular}       
\begin{description}
\item 
{
[{$^{a}$}] D and S refer to 2- and 1-sector models respectively.
Models are labelled by \tbb\ in kK.
}

\item 
{
[{$^{b}$}] `$<$' upper limit and `:' large uncertainty. Unobserved
lines: predicted intensities are given in parentheses in units of
\hb\ = 100.
}

\item 
{
[{$^{c}$}] Blends: \civ\ 4658 + \ffeiii\ 4658 + \ciii\ 4649
+ \oii\ 4651; \fsii\ 4069 + \ffev\ 4071 + \ciii\ 4069; \fariv\ 4711
+ \hei\ 4713 + \fneiv\ 4714+16.
}

\end{description}

\end{table}

\section{The central star parameters} \label{sec:more-discussion}

{
Uncertainties affecting the model geometry and the spectroscopic
constraints can enlarge the domain of acceptable solutions in the (\tbb,
$L$) plane. Uncertainties on $L$, related to $R_{{\rm in}}$ and to
\fariv/\fariii, amount to a factor of 1.6 and are therefore not too
influential in the interpretation of the stellar parameters
(Sect.~\ref{sec:Discussion}). This factor, obtained assuming a factor
1.45 uncertainty on \fariv/\fariii, allows for both observational and
theoretical uncertainties (the recombination coefficients for argon are
inaccurate, e.g., Dudziak et al.
\cite{dudziak_etal00}). The range of accessible \tbb\ is controlled by the
assumed value of $N_{{\rm in}}$(H) (Table~\ref{tab_mod_range}) and,
particularly, the intensity of \hei\,$\lambda$5876. The computed \hei\
increases rapidly towards low values of \tbb\ (alias \teff) because He/H must
increase to account for \heii\ (see model D105, Col.~2 of
Table~\ref{tab_mod_appendix}). In addition, the assumption of a strict black
body for the central star, useful in accounting for the strong \heii\ line, is
more likely to break down for lower \teff's due to the occurrence of a
discontinuity at the ionization limit of He$^+$ (e.g., Rauch \cite{rauch03}),
thus exacerbating the problem. By itself, the increase of He/H strongly
militates against too small values of \tbb. Thus, a lower limit to \tbb\ is
likely close to 120\,kK, corresponding to a minimum \teff~$\sim$~105\,kK
(Eq.~2). On the other hand, trial computations showed that dividing the present
\hei\ line intensity by 1.33, models could be obtained up to \tbb\ = 160\,kK
(D160, Col.~4 of Table~\ref{tab_mod_appendix}), equivalent to a maximum
\teff~$\sim$~145\,kK. In fact, a few models re-converged using stellar
atmosphere models (Rauch \cite{rauch03}; see Fig.~22 in P\'equignot \& Tsamis
\cite{pequignot_tsamis05}) 
showing that a more nearly correct maximum \teff\ is
almost 150\,kK. More accurate \hei\ line intensities are needed before the
range of possible \teff's can be narrowed. Note that the derived standard He/H
($=$ 0.108) is relatively large, suggesting that either the observed \heii\
line intensity is slightly overestimated or some basic model assumption is
lacking. To our knowledge, only in a situation of chemical inhomogeneity of the
PN could the derived He/H be lowered (P\'equignot et al. 2002). Indications in
favour of such a possibility are: (i) \hg\ is observed to be stronger than
expected, and (ii) the \fsii+\ciii\,$\lambda$4069 and
\civ+\ffeiii\,$\lambda$4658 blends are severely underestimated in all models
(Table~\ref{tab_mod_appendix}). These spectral regions include C and O
recombination multiplets which 
in many PNe can only be accounted for with
models that harbour H-deficient inclusions (P\'equignot et al. 2002). The
current observations of Bennu are not sufficiently deep to allow us to pursue
this possibility further.
}

\end{document}